# Three-dimensional quasi-quantized Hall insulator phase in SrSi$_2$

Kaustuv Manna[1,2*], Nitesh Kumar[1,3], Sumanta Chattopadhyay[4], Jonathan Noky[1], Mengyu Yao[1], Joonbum Park[4], Tobias Förster[4], Marc Uhlarz[4], Tirthankar Chakraborty[1], B. Valentin Schwarze[4], Jacob Hornung[4], Vladimir N. Strocov[5], Horst Borrmann[1], Chandra Shekhar[1], Yan Sun[1], Jochen Wosnitza[4,6], Claudia Felser[1], Johannes Gooth[1,6#]

[1]*Max Planck Institute for Chemical Physics of Solids, Nöthnitzer Straße 40, 01187 Dresden, Germany*

[2]*Indian Institute of Technology- Delhi, Hauz Khas, New Delhi 110 016, India*

[3] *S. N. Bose National Centre for Basic Sciences, Salt Lake City, Kolkata - 700 106, India*

[4]*Hochfeld-Magnetlabor Dresden (HLD-EMFL) and Würzburg-Dresden Cluster of Excellence ct.qmat, Helmholtz-Zentrum Dresden-Rossendorf, 01328 Dresden, Germany*

[5]*Swiss Light Source, Paul Scherrer Institut, CH-5232 Villigen PSI, Switzerland*

[6]*Institut für Festkörper- und Materialphysik, Technische Universität Dresden, 01062 Dresden, Germany*

*Kaustuv.Manna@physics.iitd.ac.in, #johannes.gooth@cpfs.mpg.de

**Abstract**

**In insulators, the longitudinal resistivity becomes infinitely large at zero temperature. For classic insulators, the Hall conductivity becomes zero at the same time. However, there are special systems, such as two-dimensional quantum Hall isolators, in which a more complex scenario is observed at high magnetic fields. Here, we report experimental evidence for a quasi-quantized Hall insulator in the quantum limit of the three-dimensional semimetal $SrSi_2$. Our measurements reveal a magnetic field-range, in which the longitudinal resistivity diverges with decreasing temperature, while the Hall conductivity approaches a quasi-quantized value that is given only by the conductance quantum and the Fermi wave vector in the field-direction. The quasi-quantized Hall insulator appears in a magnetic-field induced insulating ground state of three-dimensional materials and is deeply rooted in quantum Hall physics.**

**Main text**

The quantum Hall insulator (QHI) is one of the ground states in two-dimensional (2D) electron systems exposed to a strong magnetic field $B$. It is characterized by a diverging longitudinal resistivity ($\rho_{xx} \rightarrow \infty$) as $T \rightarrow 0$ and a finite Hall conductivity ($\sigma_{xy}$) that approaches the value $e^2/h$ that contains only fundamental constants: the elementary charge $e$ and the Planck constant $h$.[1, 2, 3, 4, 5, 6, 7] This state differs fundamentally from a classical insulator for which $\rho_{xx}$ diverges and $\sigma_{xy}$ vanishes at zero temperature.[8, 9, 10, 11, 12]

The QHI state usually occurs in between two other ground states of 2D systems. The first is the quantum Hall liquid (QHL)[13, 14], which is characterized by a vanishing $\rho_{xx}$ and a quantized $\sigma_{xy} = v \cdot e^2/h$, where $v$ is a rational number ($\rho_{xx} \rightarrow 0$ and $\sigma_{xy} \rightarrow v \cdot e^2/h$ as $T \rightarrow 0$). The second is ordinary Hall insulator (HI),[15] which is characterized by a diverging $\rho_{xx}$ and $\sigma_{xy} \approx e \cdot n/B$ where $n$ is the density of electrons in zero magnetic field ($\rho_{xx}(B) \to \infty$ and $\sigma_{xy}(B) \to e \cdot n/B$ as $T \to 0$). By changing the applied magnetic field and/or disorder, it is possible to drive transitions between the different ground states and to thereby study the quantum critical phenomena associated with them.[1, 8]

In a three-dimensional (3D) electron system, $\sigma_{xy}$ is never truly quantized.[16] Recently, however, a 3D relative of the QHL was reported as the quasi-quantized Hall effect in the Dirac semimetals ZrTe$_5$,[17, 18] HfTe$_5$[19, 20] and in doped InAs.[21] The characteristic of the 3D quasi-quantized Hall effect, i.e. the 3D quasi-quantized Hall liquid (QQHL), in the quantum limit (where only the lowest Landau band is occupied) is that $\sigma_{xy}$ is given by only the conductance quantum $e^2/h$, scaled by the Fermi wave vector $k_F$ of electrons along $B$ ($\sigma_{xy}(B) \to e^2/h \cdot k_F/\pi$ as $T \to 0$), while $\rho_{xx}$ approaches a finite, non-quantized value for $T \to 0$. The value of $\rho_{xx}$ depends on sample details, such as the residual resistivity at low temperatures. Importantly, the QHL and QQHL have a common origin: both phases are directly related to the Berry curvature of the electron states in their respective dimension.[16] This naturally raises question of whether relatives of the insulating ground states associated with the quantum-Hall regime in two dimensions might also occur in 3D systems.

In fact, a $B$-driven transition between a QQHL and 3D HI has recently been observed deep in the quantum limit of ZrTe$_5$,[17, 18] HfTe$_5$[19, 20], and doped InAs.[21] However, the transition from the QQHL to the 3D HI through a quasi-quantized Hall insulator (QQHI) phase ($\rho_{xx}(B) \to \infty$ and $\sigma_{xy}(B) \to e^2/h \cdot k_{F,z}/\pi$ as $T \to 0$), i.e., the 3D equivalent of the 2D QHI, has not yet been observed. Here we report the QQHI state in a 3D semimetal SrSi$_2$, containing near-spherical Fermi surface.

SrSi$_2$ crystalizes in the chiral space group $P4_132$ (#213) with a helical arrangement of Si atoms along the (111) direction (Fig. 1a). Originally, SrSi$_2$ was predicted to be a Weyl semimetal with a higher Chern number of ±2.[22, 23] Nevertheless, recent angle-resolved photoemission spectroscopy (ARPES) measurements reveal that SrSi$_2$ exhibits only trivial pockets with a parabolic dispersion at the Fermi energy $E_F$.[24] Ab-initio calculations of the band structure and Fermi surface (FS) are shown in Fig. 1b and c, respectively. The lack of inversion symmetry in SrSi$_2$ leads to six spin-split pairs of nearly spherical hole pockets distributed in the momentum space directions ±$k_x$, ±$k_y$ and ±$k_z$ along the high-symmetry lines Γ–X. The larger FS (red) is labelled $\beta$, whereas its spin-split counterpart, the smaller FS (blue), is labelled $\alpha$.

High quality single crystals of SrSi$_2$ were successfully grown and detailed structural as-well-as compositional analysis were performed (see Methods for details). Figure 1d shows an ARPES (see Methods for details) image of the $k_x$–$k_y$ plane taken from one of these samples. Although the spin splitting between the $\alpha$ and $\beta$ pockets is beyond the experimental resolution, the observed projection of four Fermi pockets located at ±$k_x$ and ±$k_y$ is consistent with the band structure calculations.

For the electrical transport experiments, several millimeter-size rectangular SrSi$_2$ single crystal bars were prepared such that the edges of the bars aligned along the crystallographic directions (crystals containing {100} planes or crystals with their length along [110], width along [111] and height along [-112], etc.), that correspond to the indexes $x$, $y$, and $z$ of the transport coefficients. On bars aligned in this way, we measured $\rho_{xx}$ and $\rho_{yx}$ with a standard four-probe low-frequency lock-in technique in DC and pulsed magnetic fields (see Methods for details) at various $T$. At $B = 0$, $\rho_{xx}$ and $n$ exhibit the typical behavior of an intrinsically doped semiconductor (Figs. 1e,f).[25] $\rho_{xx}$ increases with increasing $T$, while $n$ remains nearly constant up to 75 K, indicating a metallic regime. Around 75 K, however, thermal activation across the band gap sets in, causing the charge-carrier concentration $n$ to increase with increasing $T$ and $\rho_{xx}$ to decrease. d$\rho_{yx}(B)$/d$B$ is positive below 75 K, indicating hole-dominated transport in the metallic temperature-regime, which agrees with our band structure calculations and ARPES experiments. An important detail that distinguishes our study from past experiments on the quasi-quantized 3D Hall regime, is that our samples have a much lower charge-carrier mobility $\mu$ (Fig. 1f). At 2 K, we estimate $\mu = 1.21 \times 10^3$ cm$^2$V$^{-1}$s$^{-1}$, which is approximately two orders of magnitude lower than the carrier mobility for ZrTe$_5$[17, 18], HfTe$_5$[19, 20], or doped InAs.[21] This is a key difference because in 2D systems, the transition to the QHI is tied to samples with particularly low $\mu$,[2] in which incoherent scattering dominates.[8]

To characterize the FS morphology of our samples, we measured the Shubnikov-de Haas (SdH) oscillations[26] with respect to the main crystal axes at 2 K (Fig. 2a). We found two fundamental SdH frequencies (Fig. 2b) that do not change for different field alignments within the experimental resolution (Fig. 2c,d), which is consistent with two 3D spherical Fermi pockets of different sizes. Based on the ab-initio band structure calculations, we assigned the obtained frequencies $f_\alpha = 5.3$ T and $f_\beta = 12.8$ T to the $\alpha$-pocket and $\beta$-pocket, respectively. The corresponding zero-field Fermi wave vectors $k_{F,0,i}$ ($i = \alpha, \beta$) were then estimated using the Onsager relation $k_{F,0,i} = \sqrt{f_i 4\pi e/h}$, yielding $k_{F,0,\alpha} = 0.013$ Å$^{-1}$ and $k_{F,0,\beta} = 0.020$ Å$^{-1}$. Further details of the band- structure analysis can be found in Supplementary Fig. S2. We want to explicit point out two results of this analysis: the quantum limits for the $\alpha$ and $\beta$ pockets in our SrSi$_2$ samples have already been reached for fields with $B_{QL,\alpha} \approx 5.3$ T and $B_{QL,\beta} \approx 12.8$ T, respectively and the $\alpha$-pockets are already completely gapped out above $B_{G,\alpha} \approx 11.4$ T (see SI for details). Thus, above $B_{QL,\beta}$, all charge carriers are in the lowest Landau band of the $\beta$-pockets.

We now identify the different phases in SrSi$_2$ above $B_{QL,\beta}$ at low temperatures. More precisely, we measured in the configuration of the magnetic field aligned along the $z$ axis ($B_z$). Experimental identification of an insulating or metallic phase is based on extrapolation of the measured $\rho_{xx}$ in $B$ and at a finite temperature to $T = 0$. This is always an ambiguous process.

However, if $\rho_{xx}$ increases (exponentially) with decreasing $T$, the state is usually considered to be an insulator. Conversely, if $\rho_{xx}$ decreases with decreasing $T$, the state is regarded as a metal. In a 2D electron system, the transition between an insulating phase and a metallic phase in the quantum Hall regime can be characterized by a critical magnetic field value $B_c$, for which $\rho_{xx}$ is $T$-independent and where the derivative of the $T$-dependence changes sign on each side of the transition.[27] In experiments, $B_c$ is usually extracted from the intersection of $\rho_{xx}(B)$ curves measured at various values of $T$.[1] The transition between the QQHL and HI in the 3D quasi-quantized Hall regime has been characterized in a similar way in ZrTe$_5$.[17]

Following this procedure, we plot the $\rho_{xx}(B)$ and $\sigma_{xy}(B)$ of our SrSi$_2$ sample for various $T$ in Fig. 3a and b, respectively. Starting from low fields, we find that SrSi$_2$ enters the metallic 3D QQHL phase[16] just above $B_{QL,\alpha}$. In this phase, $\rho_{xx}$ decreases monotonically with decreasing $T$ (Fig. 3c), approaching a finite, non-quantized value, whereas $\sigma_{xy}$ converges to $6 \cdot e^2/h \cdot k_F/\pi$ for $T \to 0$ (Fig. 3b,d). The six-time degeneracy of $\sigma_{xy}$ comes from the six $\beta$-pockets in the Brillouin zone (BZ). For 3D systems with a parabolic band structure, the $k_F$ of the lowest Landau band is $B$-dependent.[16, 21] In particular, $k_F = \sqrt{k_{F,0}^2 - 2eB/h}$. Consequently, in SrSi$_2$ the quasi-quantized Hall conductivity above $B_{QL,\beta}$ in the zero-$T$ limit is given by $\sigma_{xy} = \frac{6 \cdot e^2}{h} \sqrt{k_{F,0,\beta}^2 - 2eB/h}/\pi$. With increasing $B$, a transition at $B_c \approx 20.2$ T from the QQHL to an insulating phase can be identified in the intersection of the $\rho_{xx}$ curves obtained at different values of $T$ in Fig. 3a. In the insulating phase (Fig. 3d), a striking observation can be made: just above $B_c$, there is a field range of approximately 2 T in which $\sigma_{xy}$ remains quasi-quantized. In exact terms, between $B_c$ and approximately $B_T = 22.2$ T, the deviation of $\sigma_{xy}$ from $6 \cdot e^2/h \cdot k_{F,\beta}/\pi$ at 0.6 K is less than 1%, even though $\rho_{xx}(B)$ increases monotonically for $T \to 0$. This reflects the expected features of a magnetic field-tuned 3D QQHI phase.

Theoretically, the magnetic field-driven transitions from a liquid phase to an insulating phase in the quantum Hall or quasi-quantized Hall regime are considered transitions between delocalized and localized phases.[8, 9, 10, 11, 12] They can be understood in the context of percolation path network models, which provide similar results in two[8, 9, 10, 11, 12] and three[28, 29] dimensions. The insulating phases are not entirely insulating in their bulk but are rather described as separate clusters of liquid states that exist in potential minima induced by random impurities and defects throughout the samples. The transport is then obtained via a scattering matrix, which parametrizes the tunneling of electrons (holes) from one liquid cluster to the other, leading to the diverging $\rho_{xx}$ for $T \rightarrow 0$. When the phase coherence of the electron (hole) wavefunctions is maintained after each tunneling event, the insulating states are ordinary HI.[28, 30, 31] Conversely, for 2D systems it has been shown that if the phase coherence of the electron (hole) wavefunctions is destroyed after each tunneling event, $\sigma_{xy}$ is quantized; *i.e.*, the system is a QHI.[30] A similar mechanism may distinguish the ordinary HI from the QQHI phase in 3D systems, where a perpendicular magnetic field can change the size of the liquid clusters and gradually merge them together. The field $B_C$, at which a trajectory of the liquid phase percolates through the entire system, accordingly marks the transition into the QHL or QQHL phase in two or three dimensions, respectively.

The isotherms of $\rho_{xx}$ for such magnetic field-induced metal-insulator transitions typically obey a universal scaling law with the parameter $(B - B_c)T^{-1/\zeta}$, where $\zeta$ is the critical exponent. This applies to transitions between the QHL and QHI in 2D systems[27] as well as to transitions between the 3D QQHL and the ordinary HI[28, 30], as in ZrTe$_5$[17]. In Fig. 3e, we show such a scaling analysis for SrSi$_2$ above $B_{QL, \beta}$, and, indeed find that all isotherms of $\rho_{xx}$ fall on a single curve as a function of $(B - B_c)T^{-1/\zeta}$ for $\zeta \approx 9.9$.

Deeper in the insulating state, we observe that $\sigma_{xy}$ begins to deviate from its quasi-quantized value and the state seems to evolve continuously to a 3D HI. The crossover between these insulating states is defined by a transition field $B_T$, at which the experimentally determined $\sigma_{xy}$ deviates more than 1 % from $6 \cdot e^2/h \cdot k_{F,\beta}/\pi$. The crossover can be understood as a field-driven change from a state of constant Fermi level $E_F$ to a state with a constant charge-carrier density.[18,19,21] The theoretical considerations presented above strongly depend on $E_F$ being fixed, which ensures that the lowest Landau bands shift linearly with $B$. In materials with relatively small Fermi areas, such as $SrSi_2$, such a situation can occur as a result of defect states that absorb some of the conduction electrons (holes), thereby ensuring overall charge neutrality. Nevertheless, the number of these defect states in real samples is limited. Moreover as soon as the defect states are fully occupied, $E_F$ begins to move with $B$ to avoid large charging energies and $\sigma_{xy}$ approaches its classical value. In agreement with our observations, such a crossover usually occurs in large fields as the bottom of the last populated Landau band approaches $E_F$. In Fig. 3b, the hypothetical field, where $SrSi_2$ enters the bandgap under the assumption of a fixed $E_F$ is identified as $B_{G,\beta}$.

A specific feature of the quasi-quantized Hall regime in 3D systems is that $B_{QL}$ and $\sigma_{xy}$ sensitively depend on the size of the FS in a zero field, *i.e.*, $k_{F,0,\beta}$.[16,18,19,21] To test whether our observations in particular, the onset and $\sigma_{xy}$ in the 3D QQHI state scale accordingly, we grew an additional batch of $SrSi_2$ samples with 2 % Ca doping, which is known to shift the intrinsic $E_F$ closer to the valence band edge.[24] The details of the experiments and analysis of the Ca-doped $SrSi_2$ samples are provided in Supplementary Figs. S3, S4 and S5. The results observed for the undoped sample are fully reproduced with Ca doping, but they are scaled by a smaller $k_{F,0,\beta}$. This is an important cross check that reaffirms the occurrence of the 3D QQHI state in $SrSi_2$.

In conclusion, we have reported experimental evidence for a quasi-quantized Hall insulator in the lowest Landau band (quantum limit) of the 3D semimetal SrSi$_2$. Our measurements reveal a magnetic field range, in which the longitudinal resistivity diverges with decreasing temperature and the Hall conductivity only depend on the conductance quantum and the Fermi wave vector in the field-direction ($\rho_{xx}(B) \to \infty$ and $\sigma_{xy}(B) \to 6 \cdot e^2/h \cdot k_{F,\beta}/\pi$ as $T \to 0$). The quasi-quantized Hall insulator is a magnetic field induced insulating ground state of a 3D material that is deeply rooted in quantum physics.

## Methods

*SrSi$_2$ Crystal Growth and Structural Refinement*

High quality single crystals of SrSi$_2$ were grown from the stoichiometric polycrystalline ingots. Typical size of the grown SrSi$_2$ crystals were ~5 mm diameter and upto 30 mm length. The single-phase crystallinity was checked with a white beam backscattering Laue X-ray photography at room temperature. A single, sharp Laue spots clearly revealed the excellent quality of the grown crystals without any twining or domains. A representative Laue pattern of an annealed crystal, superposed with a theoretically simulated one is shown in Fig. S-1. The chemical composition of the grown crystals was verified by energy-dispersive X-ray (EDX) spectroscopy, the results of which matched the target stoichiometric composition of SrSi$_2$. Rigorous single crystal X-ray diffraction experiments were performed to analyze the structural chirality of the grown crystals. The results are discussed in the SI. The refined crystal structure confirms of $P4_132$ (No.213) type, with Flack parameter ~ 0.01(5), which is well in accordance with a single-handed chiral domain.

*Angle-resolved photoemission spectroscopy: ARPES*

The ARPES experiments were conducted using soft x-ray ARPES (SX-ARPES) at the ADRESS beam-line of the Swiss Light Source using a SPECS analyzer.[32, 33] SrSi$_2$ single crystals of (001) orientation and typical size ~ 4×2×2 mm$^3$ were cleaved in-situ at 16 K in a high-vacuum chamber (with a base vacuum better than 5×10$^{-11}$ torr).

*First-principles calculations*

The calculations of the electronic structure were performed using density-functional theory (DFT) as implemented in VASP utilizing an augmented plane-wave basis set and pseudopotentials.[34] The exchange-correlation potential was evaluated with the hybrid HSE approximation.[35] From the DFT eigenvalues and eigenfunctions, maximally-localized Wannier functions (MLWFs) were extracted using the package Wannier90[36] and in the following a tight-binding Hamiltonian was generated. With the Hamiltonian, the Fermi surface was evaluated on a mesh of $301 \times 301 \times 301$ points in the Brillouin zone.

*Electrical and thermoelectric transport measurements*

Low magnetic field resistivity measurements upto 14 T were performed in commercial physical property measurement systems (PPMS-9T, and PPMS-14T Quantum Design) using the Electrical Transport Option (ETO) and the AC Transport (ACT) modes with the rotator option. The longitudinal and Hall resistivity measurements were performed using a 4-wire geometry with silver paint and 25-µm platinum wires. To minimize self-heating, a low current of 0.2 mA was maintained. The pulse magnetic field transport measurements were conducted at the Dresden High Magnetic Field Laboratory, Germany using 4-point resistivity measurements with a current of 0.2 mA. A 70 T non-destructive pulsed magnet driven by a capacitor bank with a 150 ms pulse duration and equipped with a $^3$He cryostat insert was used for the measurements.

## Notes

The authors declare no competing financial interests.

## Author contributions

K.M started the project. K.M, C.F and J.G. supervised the overall progress. The single crystals were grown by K.M. The basic structural characterization and chemical property analysis of the bulk crystals were carried out by K.M and H.B. with the help of N.K. and C.S. Final transport devices were fabricated by K.M., N.K. and C.S. Transport experiments and analysis were carried out by K.M. and S.C. with the help of N.K., C.S., J.P., T.F., M.U., T.C., B.V.S., J.H and J.W.. ARPES measurements were performed by M.Y. with the help of V.N.S.. J.N. and Y.S provided the Theoretical model of the three-dimensional quantum Hall effect and the ab-initio calculations. K.M., N.K., J.N. and J.G. analyzed the data and wrote the manuscript. All authors contributed to the interpretation of the data and discussion of the scientific content of the manuscript.

## Acknowledgments

K.M., N.K. and C.F. acknowledges financial support by European Research Council (ERC) Advanced Grant No. 742068 ("TOPMAT"). C.F. and J.W. acknowledges support by Deutsche Forschungsgemeinschaft (DFG) under SFB 1143 (Project No. 247310070) and Würzburg-Dresden Cluster of Excellence on Complexity and Topology in Quantum Matter-ct.qmat (EXC 2147, project no. 39085490). J.W. also acknowledges support from the DFG through the ANR-DFG grant Fermi-NESt, and by Hochfeld-Magnetlabor Dresden (HLD) at HZDR, member of the European Magnetic Field Laboratory (EMFL). J.G. acknowledges support from the European Union's Horizon 2020 research and innovation program under Grant Agreement ID 829044 "SCHINES". Additionally, K.M. acknowledges Max Plank Society for the funding support under Max Plank–India partner group project.

## Supplementary Information

The Supplementary Information contains Supplementary Sections on "Structural characterization of $SrSi_2$ single crystal", "Charge-carrier density and mobility from Hall measurements", "Mapping of the Fermi surface by analyzing Shubnikov-de Haas oscillations", "The special magnetic field values", Supplementary Fig. S1 – S6, Supplementary Tables S1 and S2.

**MAIN FIGURES**

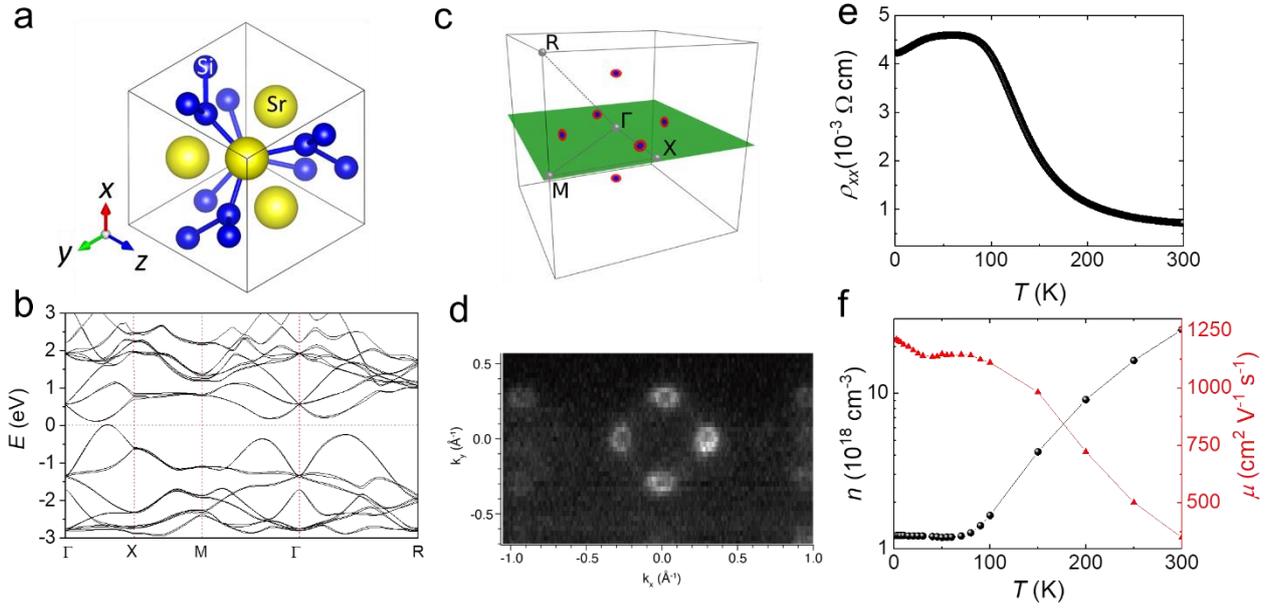

**Figure 1 | Electronic structure and electrical characterization of chiral SrSi$_2$: a**, Crystal structure of SrSi$_2$ with chiral space group $P4_132$ (#213). **b**, Trivial band structure of SrSi$_2$ in agreement with a recent ARPES study.[24] **c**, Six pairs of spin-split hole Fermi pockets in the BZ along $\pm k_x$, $\pm k_y$ and $\pm k_z$. **d**, Constant energy surface at $E_F$ and $k_z = 0$ measured by ARPES acquired with a photon energy $h\nu = 342$ eV. Cross-section of a set of four near-spherical Fermi pockets lying in the $k_x$–$k_y$ plane is clearly visible. The corresponding plane is highlighted in green in (c). **e**, Temperature dependent resistivity of SrSi$_2$ with current passed along [100] in the absence of a magnetic field. **f**, Temperature dependent charge-carrier concentration ($n$) and mobility ($\mu$) of SrSi$_2$ estimated using the low magnetic field slope of $\rho_{yx}(B)$.

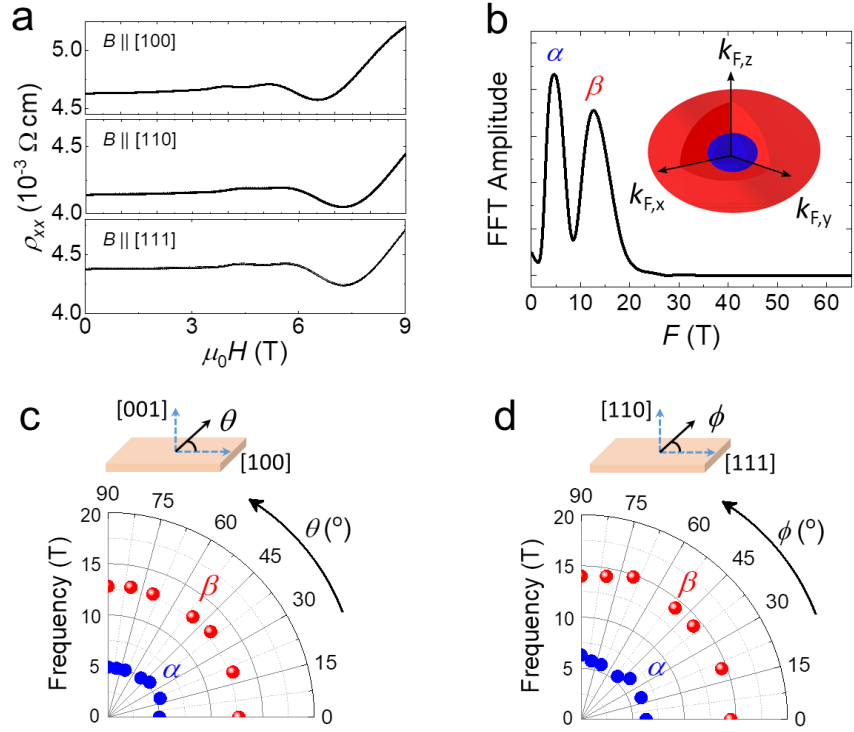

**Figure 2 | Near-spherical Fermi surface of SrSi$_2$. a**, Shubnikov-de Haas (SdH) oscillations at 2 K with a magnetic field applied along important cubic crystallographic directions: [100], [110], and [111], which exhibit nearly identical quantum oscillations. **b**, FFT amplitudes for SrSi$_2$ determined from SdH oscillations for *B* along [100]. The inset shows the near-spherical Fermi-surface pockets predicted by the calculations. **c-d**, Angle dependent SdH oscillations with, identical oscillation frequencies, that are in agreement with the nearly spherical Fermi-surface pockets in SrSi$_2$ are revealed.

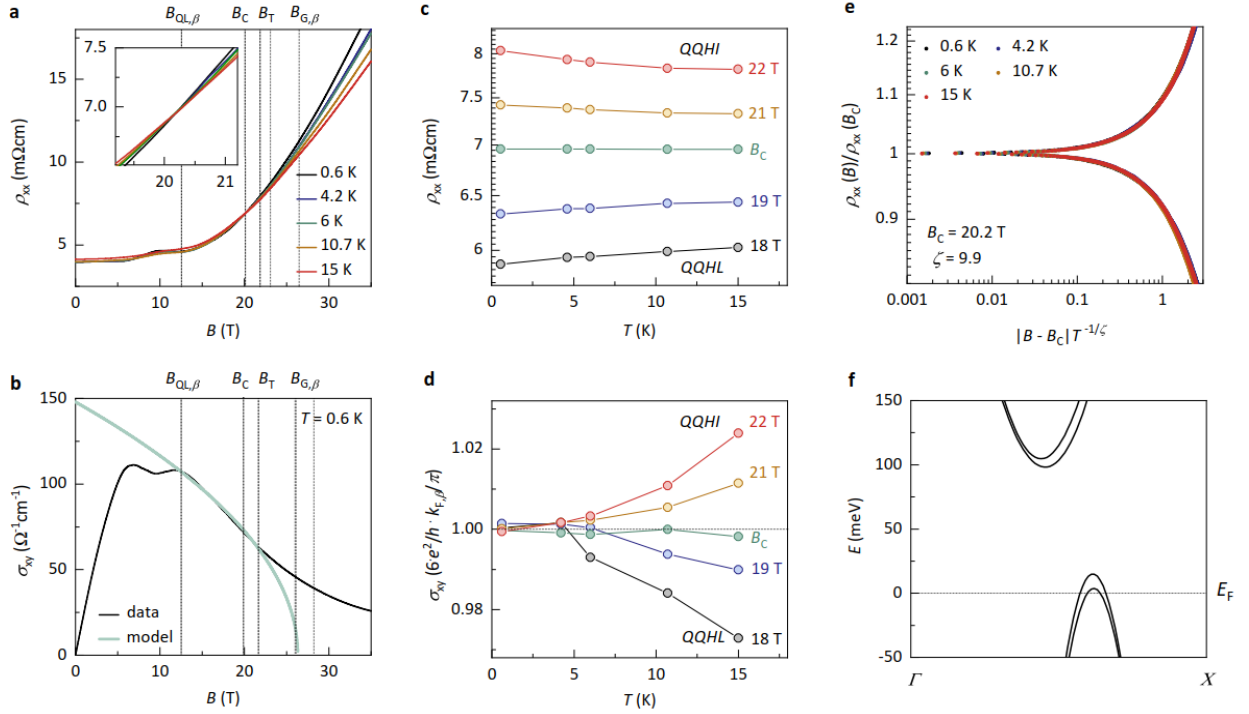

**Figure 3 | Three-dimensional (3D) quasi-quantized Hall insulator (QQHI) in the quantum limit of undoped SrSi$_2$. a,** Longitudinal resistivity $\rho_{xx}$ as a function of the magnetic field $B$ for various temperatures $T$. The field at which the $\beta$-pocket enters the quantum limit (the lowest Landau band) is denoted $B_{QL,\beta}$. $B_c$ marks the transition point from a metallic to an insulating state. **b**, Hall conductivity $\sigma_{xy}$ as a function of $B$ at $T = 0.6$ K. In the insulating phase, between $B_c$ and $B_T$, the experimentally determined $\sigma_{xy}$ (black curve) scales with the calculated 3D quasi-quantized Hall conductivity value (light green curve) estimated using the experimentally extracted Fermi wave vector $k_F$ along $B$, the electron charge $e$ and the Planck constant $h$. $B_{G,\beta}$ marks the theoretical field at which undoped SrSi$_2$ hypothetically enters the band gap. **c**, $\rho_{xx}$ as a function of $T$ for various $B$ around $B_c$. **d**, $\sigma_{xy}$ as a function of $T$ for various $B$ around $B_c$,. **e**, Normalized resistivity $\rho_{xx}(B)/\rho_{xx}(B_c)$ as a function of the scaling parameter $|B − B_c|T^{−1/\zeta}$, with the critical exponent $\zeta = 9.9$. **f**, Zoomed-in view of the band structure of undoped SrSi$_2$, revealing two hole pockets.

# Supplementary Information

**Table of Content**

**S1: Structural characterization of SrSi$_2$ single crystal:**

The single crystal diffraction experiments on the SrSi$_2$ single crystals were performed on a Rigaku AFC7 four-circle diffractometer with a Saturn 724+ CCD-detector applying graphite-monochromatic Mo-Kα radiation. For the measurements, small crystal pieces were broken from the large single crystal block. The small fragments were mounted on Kapton loops for the data collection. The crystal refinement results are summarized in Table-S1. The complete crystallographic information has been deposited and is available citing CSD-xxxxxx.

**S2: Charge-carrier density and mobility from Hall measurements**

The charge-carrier concentration ($n$) and mobility ($\mu$) are estimated from the magnetic field slope of $\rho_{yx}(B)$ in the range ±4 T using $n = 1/e.R_0$ and $\mu = R_0/\rho_{xx}$, respectively. Here $R_0$ is the Hall coefficient, calculated from the slope of $\rho_{yx}(B)$. The temperature dependence of $n$ and $\mu$ are plotted in Fig. 1f.

## S3: Fermi surface mapping from temperature dependent Shubnikov-de Haas oscillations

We mapped the Fermi surface of the SrSi$_2$ single crystals by analyzing the Shubnikov-de Haas oscillations. The temperature ($T$)-dependent longitudinal electrical resistivity $\rho_{xx}(B)$ was used for the analysis. Due to the near spherical nature of the Fermi-surface pockets, we restrict the discussion to the data, where the magnetic field was applied along the [001] direction. The results for both the undoped and Ca-doped SrSi$_2$ crystals are summarized in Supplementary Table S2. Fig. S2 displays the result for the undoped SrSi$_2$ single crystal. The two fundamental frequencies, $\alpha \sim 5.3$ T and $\beta \sim 12.8$ T can be clearly resolved using the fast Fourier transform (FFT) (Figs. S2a-b). The extremal cross-sectional areas ($A_F$) of the FSs are estimated from the quantum oscillation frequency ($F$) using the Onsager relation: $F = (\Phi_0/2\pi^2)A_F$, where $\Phi_0$ is the fundamental flux quantum $= 2.068 \times 10^{-15}$ Wb. $A_F$ calculated for the frequencies $\alpha$ and $\beta$ are 0.0005 and 0.0012 Å$^{-2}$ respectively, revealing the small size of the Fermi-surface pockets. The effective mass ($m^*$) is estimated from the temperature dependent SdH amplitudes (Fig. S2b) using the temperature-dependent part of the Lifshitz–Kosevich (LK) formula: $R_T = X/\sinh(X)$, where $X = 14.69\, m^*T/B$, $R_T$ is the thermal damping factor related to the temperature induced broadening of Landau levels in the Fermi-Dirac distribution, $B$ is the average field. $m^*$ for the holes in the $\alpha$ and $\beta$ pockets are found to be 0.154 $m_0$ and 0.171 $m_0$ respectively (Fig. S2c). Apart from R$_T$, the Dingle damping factor R$_D$ describes the exponential decrease of the oscillation amplitude with decreasing magnetic field. The Dingle temperature $T_D$ can be estimated from the slope of the Dingle plot (Fig. S2d) which is inversely proportional to the quantum scattering lifetime ($\tau_q$) via $\tau_q = \hbar/(2\pi k_B T_D)$. Our estimated value of T$_D$ and $\tau_q$ are 8.7 K and 1.4 x 10$^{-13}$ s, respectively. Similar in-depth analysis was also performed for the Ca-doped sample, Sr$_{0.98}$Ca$_{0.02}$Si$_2$, and the corresponding plots are shown in Fig. S3.

**S4: The special magnetic field values explained in main text**

There are four different special magnetic field values for the $\beta$ pockets: $B_{QL}$, $B_C$, $B_T$, and $B_G$. Here, $B_{QL}$ describes the field, at which the pockets reach the quantum limit, i.e. the field at which all electrons remain in the lowest Landau level. $B_C$ gives the critical field value, at which the system undergoes a phase transition from the QQHL state into an insulating phase. $B_T$ describes the threshold field, at which the results start to differ from the theoretical single-particle model with a fixed Fermi energy. Up to this field, the charge carriers redistribute themselves within the sample in a way, that the Fermi level is kept constant with the help of defect and impurity states. Above this field, this becomes energetically too costly and the Fermi level starts shifting. At last, $B_G$ gives the value, at which the last Landau level in the theoretical model leaves the Fermi energy and the system becomes fully gapped. As this state would need too much energy, it is not reached in the real sample, instead, the Fermi level starts to shift after $B_T$. Since it is not possible to experimentally measure the field $B_G$, it is evaluated by using the equation $k_F = \sqrt{k_{F,0}^2 - 2\,eB/h}$ and setting it to zero, resulting in $B_G = h\,k_{F,0}^2/(2e)$.

**Supplementary Fig. S1:**

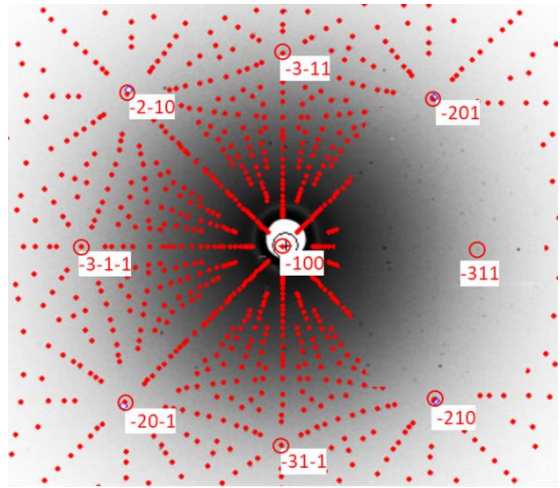

**Supplementary Fig. S1.** A representative Laue diffraction pattern of the [100] oriented SrSi$_2$ single crystal, superposed with a theoretically simulated one confirming high crystalline quality**.**

**Supplementary Fig. S2.**

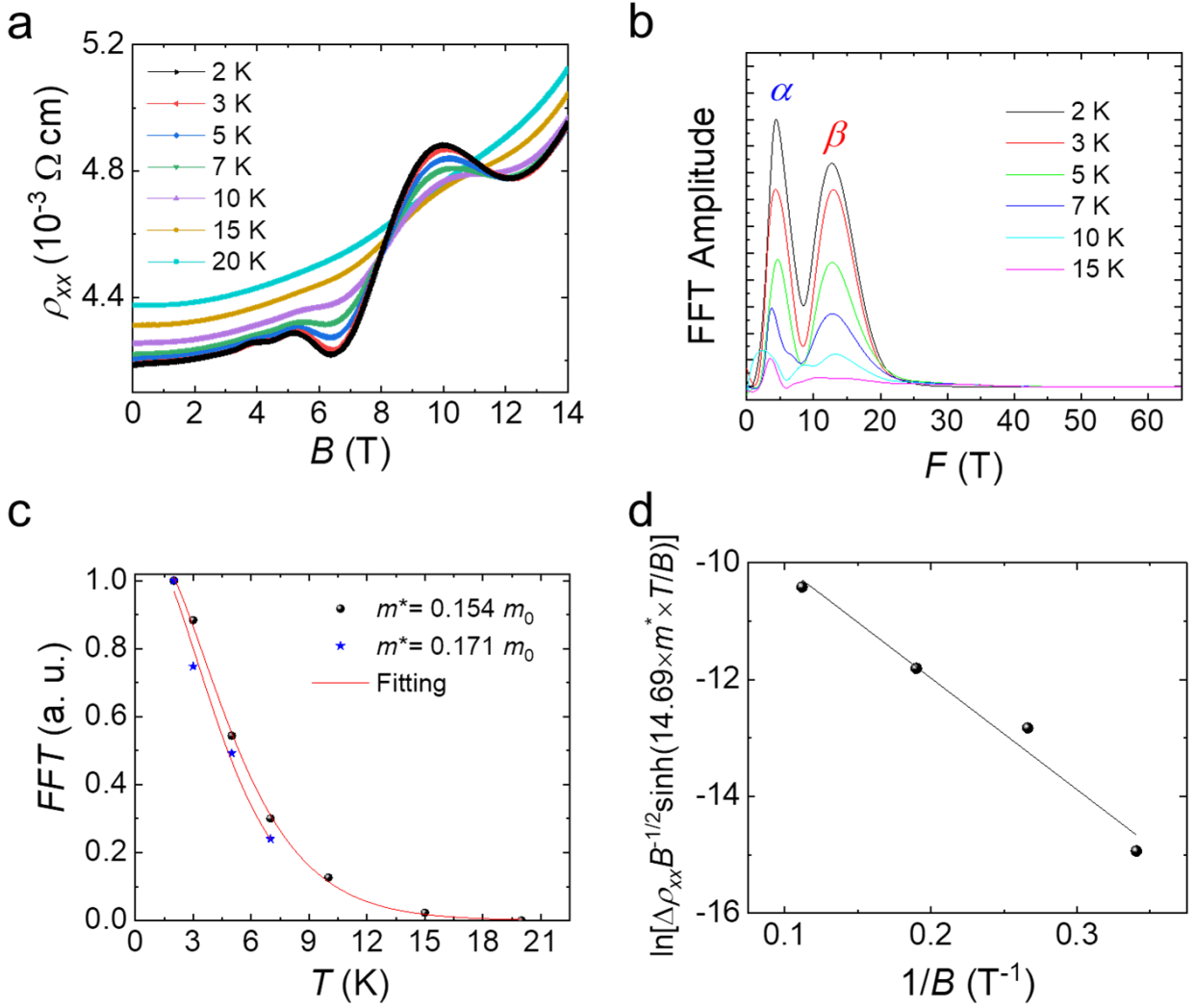

**Supplementary Fig. S2. a,** Temperature dependent longitudinal resistivity $\rho_{xx}(B)$ of SrSi$_2$ measured up to 14 T. **b,** FFT amplitudes for SrSi$_2$ determined from the SdH oscillations at various temperatures. Corresponding frequencies are indexed as $\alpha$ and $\beta$ for ~ 5.3 and 12.8 T, respectively. **c,** Fitting of the temperature dependent quantum oscillation amplitudes with the LK formula for the $\alpha$ and $\beta$ pockets to estimate the value of $m*$. **d,** Dingle plot for SrSi$_2$ from the FFT amplitude by selecting corresponding field windows of constant width in $1/B$.

**Supplementary Fig. S3.**

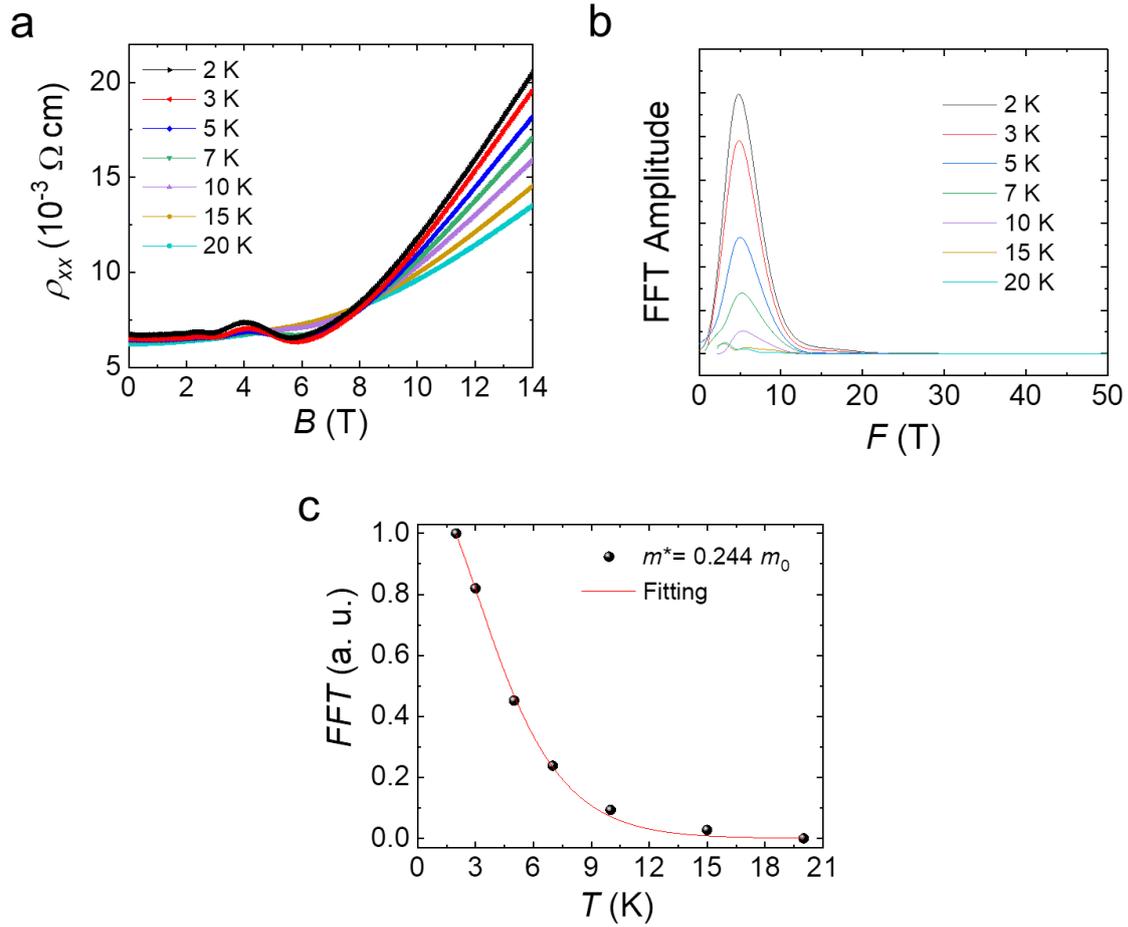

**Supplementary Fig. S3. a,** Temperature dependent longitudinal resistivity $\rho_{xx}(B)$ of the Ca-doped sample $Sr_{0.98}Ca_{0.02}Si_2$ measured up to 14 T. **b,** FFT amplitudes determined from the temperature dependent SdH oscillations, revealing one frequency at about 5.7 T. **c,** LK formula fitting of the temperature dependent quantum oscillation to estimate the value of $m*$.

**Supplementary Fig. S4.**

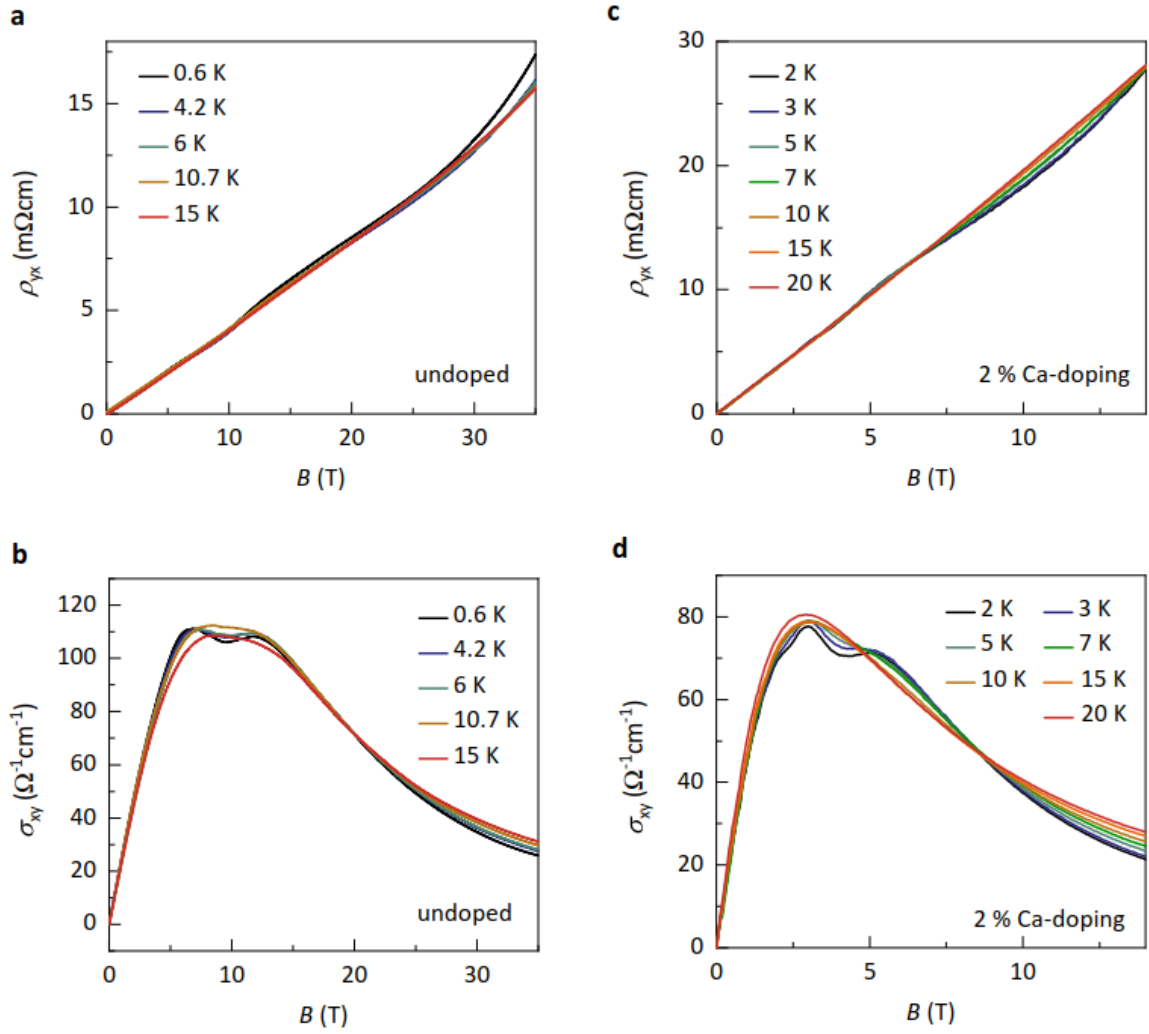

**Supplementary Fig. S4.** Temperature dependent Hall resistivity $\rho_{yx}(B)$ and the calculated Hall conductivity $\sigma_{xy}(B)$ of the undoped SrSi$_2$ (**a, b**, respectively) and the Ca-doped Sr$_{0.98}$Ca$_{0.02}$Si$_2$ (**c, d** respectively) single crystals.

**Supplementary Fig. S5.**

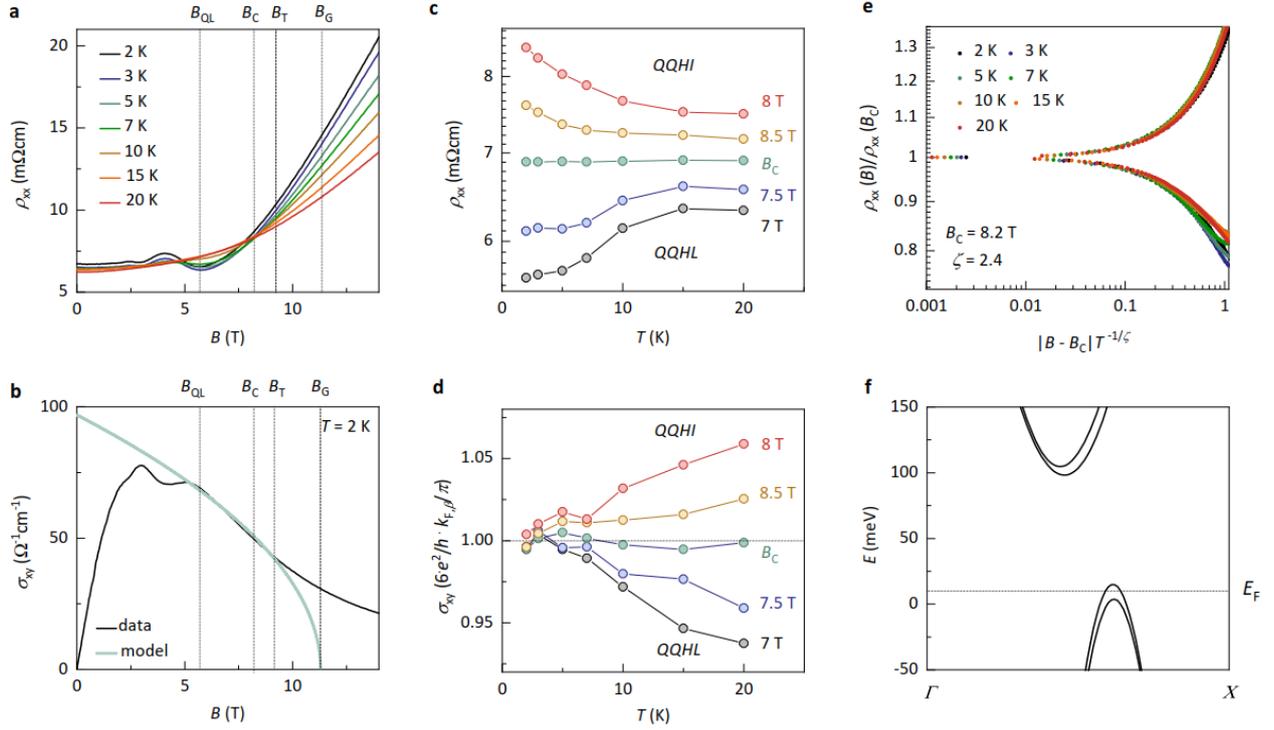

**Supplementary Fig. S5: a,** Temperature variation of the longitudinal resistivity $\rho_{xx}$ as a function of the magnetic field. The special magnetic field notations, $B_{QL}$, $B_c$, $B_T$ and $B_G$ carries the same meaning as in Fig. 3; a more detailed explanation is provided in the SI. **b,** Field dependent Hall conductivity $\sigma_{xy}$ at 2 K: in the insulating phase between $B_c$ and the transition field $B_T$, the experimentally determined $\sigma_{xy}$ (black curve) scales with the calculated 3D quasi-quantized Hall conductivity value (light green curve) estimated using the experimentally determined Fermi wave vector $k_F$ along $B$, the electron charge $e$, and the Planck constant $h$. Beyond $B_{G,\beta}$ $Sr_{0.98}Ca_{0.02}Si_2$ hypothetically enters the band gap. **c,** $\rho_{xx}$ as a function of temperature around $B_c$. **d,** $\sigma_{xy}$ as a function of $T$ for various $B$ around $B_c$, in the 2 % Ca-doped SrSi$_2$ crystal. **e,** Normalized resistivity $\rho_{xx}(B)/\rho_{xx}(B_c)$ as a function of the scaling parameter $|B - B_c|T^{-1/\zeta}$, with the critical exponent $\zeta$. **f,** Single hole pocket formation in a Ca-doped SrSi$_2$ crystal due to the shifted position of $E_F$.

**Supplementary Fig. S6.**

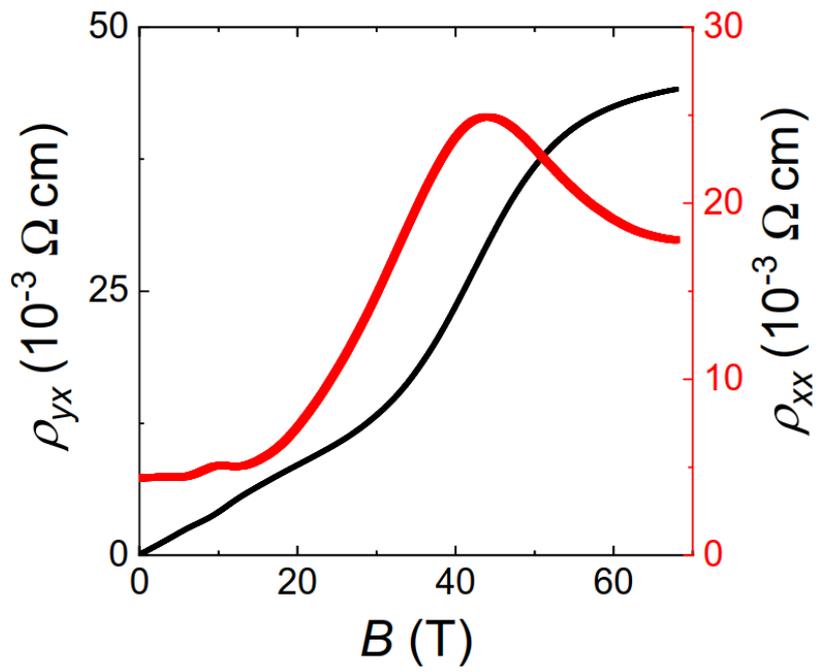

**Supplementary Fig. S6.** Field dependent longitudinal resistivity $\rho_{xx}(B)$ and Hall resistivity $\rho_{yx}(B)$ of the SrSi$_2$ single crystal measured upto 68 T using a pulse-field magnet.

**Supplementary Tables-S1: Room temperature single crystal Refinement result for SrSi$_2$.**

| Formula | SrSi$_2$ |
|---|---|
| F.W. (g/mol); | 143.8 |
| Space group; $Z$ | $P4_132$ (No.213); 4 |
| $a$ (Å) | 6.5361(4) |
| $V$ (Å$^3$) | 279.23(5) |
| Absorption Correction | Multi-scan |
| Extinction Coefficient | 0.008(3) |
| θ range (deg) | 4.4-34.4 |
| No. reflections; $R_{int}$ | 2105; 0.0222 |
| No. independent reflections | 202 |
| No. parameters | 7 |
| $R_1$; $wR_2$ (all $I$) | 0.0117; 0.0176 |
| Flack parameter x | -0.008(10) |
| Goodness of fit | 0.72 |
| Fourier difference peak and hole (e$^-$/Å$^3$) | 0.258; -0.234 |

**Supplementary Table S2. Band-structure parameters of the undoped and doped SrSi$_2$ single crystals obtained from Shubnikov-de Haas oscillations.**

| Sample | SdH frequency $B_F$ (T) | Fermi area $A_F$ ($10^{-4}$ Å$^{-2}$) | Fermi wave vector $k_F$ ($10^{-3}$ Å$^{-1}$) | Fermi wave length $\lambda_F$ (nm) | Cyclotron mass $m_c$ ($m_0$) | Fermi velocity $v_F$ ($10^5$ m/s) | Dingle temperature $T_D$ (K) | Lifetime $\tau_q$ (ps) |
|---|---|---|---|---|---|---|---|---|
| Undoped SrSi$_2$ | 5.3 | 5.05 | 12.7 | 49.5 | 0.154 | 0.95 | 8.7 | 0.14 |
|  | 12.8 | 12.2 | 19.7 | 31.9 | 0.171 | 1.33 |  |  |
| Doped SrSi$_2$ | 5.7 | 5.44 | 13.2 | 47.76 | 0.244 | 0.625 |  |  |